# THE ASTRONOMY GENEALOGY PROJECT IS TEN YEARS OLD: HERE ARE TEN WAYS YOU CAN USE IT

Joseph S. Tenn

*Department of Physics & Astronomy, Sonoma State University, Rohnert Park, CA 94928, USA.*
E-mail: tenn@sonoma.edu

**Abstract:** The Astronomical Genealogy Project (AstroGen) has been underway since January 2013. This project of the Historical Astronomy Division (HAD) of the American Astronomical Society (AAS) has been online since July 2020, courtesy of the AAS. The volunteers of the AstroGen team have systematically searched online directories, mostly at individual university libraries, for astronomy-related doctoral theses equivalent to the modern, research-based Ph.D. We now claim to be 'nearly complete' for 38 countries, although some have not been updated for a year or two or three.

The website contains a page for each astronomer and advisor, with links to the persons, universities, institutes, and the theses themselves. More than two-thirds of the theses are online in full, although some require access to a library with a subscription. There is information about nearly 37,000 individuals who have earned astronomy-related doctorates and another 5400 who have supervised them, but may not have earned such degrees themselves. Most of the latter have not yet been evaluated, but probably a majority earned doctorates in other fields, such as physics or geology.

We present some of the results of our research and discuss ten ways the reader might make use of the project.

## 1 WHAT IS ASTROGEN?

The Astronomy Genealogy Project (AstroGen) is a database of people who have earned doctorates with astronomy-related theses. Founded as a project of the Historical Astronomy Division (HAD) of the American Astronomical Society (AAS) in January 2013, it has been on line at https://astrogen.aas.org, courtesy of the AAS, since 25 July 2020. That date was chosen because it was the 159th anniversary of the awarding of the first three Ph.D.s in the United States, one of them in astronomy. A preliminary account of AstroGen (Tenn, 2016) appeared in this journal seven years ago.

We try to go back to the beginnings of the modern Ph.D., for which the thesis is supposed to be an original contribution to knowledge. We believe that AstroGen is 'nearly complete' for astronomy-related doctorates from 38 countries (and other polities). All data in this paper are as of 11 July 2023.

A challenge was deciding which theses should be in our database. Our criteria are stated in some detail on the website in the FAQs section. Briefly, we include theses that deal with the scientific study of anything that is or comes from outside the Earth, and the development of tools to facilitate such study.

AstroGen is also a database of the universities that have granted astronomy-related doctorates, and of the institutes where thesis research was done. Each astronomer has a page with his or her name, other names used professionally, Orcid or ISNI number, highest degree(s), university that awarded the highest degree, and, if applicable, the institute (not connected with the university) where the research was performed. If the highest degree was an astronomy-related doctorate, then the page also contains the title of the thesis, the name of the thesis advisor(s), and any mentors, defined as unofficial thesis advisors. The page includes links to as many of these as possible. The page also includes a family tree, showing the astronomer's academic children (students) and ancestors (advisors, their advisors …). Note that we use the American term *advisor* for the person who directs the thesis research. Equivalent terms include supervisor, guide, *directeur*, *Betreuer*, *promotor*, and others. An example of such a page (without the family tree) is shown in Figure 1 (all figures not otherwise credited are by the author, as are all tables).

The name of the university that awarded the degree is given, in English, as it was at the time of the degree. This name is linked to a page giving the name in the local language as well as the names of successor institutions up to the present, with the current name linked to the current university website. For our purposes, we define a university as any institution that awards astronomy-related doctorates, even though some are not genuine universities. See Figure 2 for an example of a university page for an institute that awards doctorates.

Individuals who have not earned doctorates with astronomy-related theses are included in AstroGen if they have directed research for such theses. Nowadays, most of these have earned what we call 'Other Doctorates', mostly in physics, but in some cases in geology, computer science, meteorology, mathematics, or engineering. A few people with master's degrees, bachelor's degrees, or even no degree have also supervised astronomy-related doctorates. The last person who had no earned degree but supervised such a doctorate was E. Arthur Milne (Figure 3) in 1940. For such persons, we include only the personal information, the highest degree (e.g., Ph.D.), the university, and the year. We do not list the thesis title or advisor(s), and we do not go back to any earlier generations.

Figure 1: A typical astronomer page. Note that it has links to the person, the University that awarded their highest degree, their advisor's page, and the student's name and university. There are also links to forms to submit additions or corrections to AstroGen. On the website, the page is accompanied by a family tree showing the astronomer's students (but not their students) and all of their known academic ancestors back to the first ones who did not earn astronomy-related doctorates.

Figure 2: A typical university page. In this case it is not truly a university, but we treat it as one since it awards doctoral degrees.

There is much more about the project in the FAQs on the AstroGen website.

## 2 WHAT HAVE WE ACCOMPLISHED?
### 2.1 Persons in AstroGen

Table 1 shows who is currently listed in the AstroGen database. Those who have not earned astronomy-related doctorates are included because they supervised research for such degrees. Those who earned two astronomy-related doctorates or two other doctorates were counted only once in the first two rows. The subtraction later is for those who had one of each. The large number of advisors whose highest degrees are unknown reflects the shortage of volunteers to seek such information online.

### 2.2 Theses in AstroGen

Table 2 provides information about the astronomy-related theses currently in our database. We have recorded more than 37,400 such theses, and readers may be surprised to learn that 89% are in English. This will be discussed below in Section 4.7. Another potential surprise is that more than two-thirds of the theses are fully online, although not all are available to all readers. The theses listed as being 'on ProQuest' are available in full to anyone who has privileges at a library that subscribes to this proprietary database. Fortunately, a great many university libraries subscribe.

## 3 HOW DO WE DO IT?

Nearly all our data have been obtained by searching the World Wide Web. ProQuest is a major source, as it provides metadata for far more theses than it makes available online. A successor to University Microfilms, which was the depository of record for theses in the United States and other countries for decades, ProQuest now lists more than three million doctoral dissertations, including 24,540 that result from a search on 'astronomy OR astrophysics OR cosmology OR planetary science'.

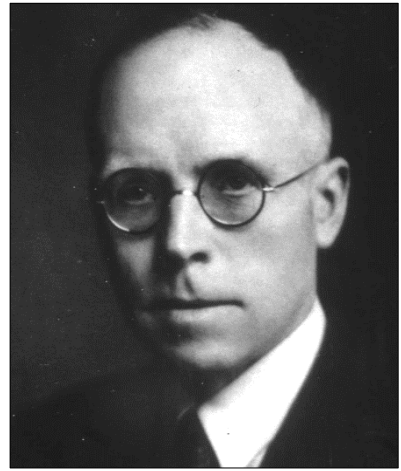

Figure 3: E. Arthur Milne (1896–1950) was the last person to supervise an astronomy-related doctorate without having earned any degree himself. He supervised at least six, from 1931 to 1940, as the Rouse Ball Professor of Mathematics at the University of Oxford. His higher education ended after 1.5 years due to World War I, and after the War, he immediately started his illustrious career as a theoretical astrophysicist (1945 Photograph © Royal Astronomical Society, courtesy: Dr Sian Prosser).

There are a few international directories, such as DART, which links to more than 1.3 million theses from 580 universities in 29 European countries, and nationwide directories in several countries, including Brazil, Canada, France, Greece, Spain, Sweden, and the United Kingdom. (Some of these only include thes-es from recent years, when the theses were 'born digital'.) However, most of our searching is done on the websites of university libraries. Some of these are easy to navigate and search; others are more challenging and may be incomplete. Even though a great many theses are now in English, it helps to have some familiarity with the language of the country to navigate the library sites. A small group of volunteers has gathered all the information currently in the database.

Table 1: Persons in AstroGen.

| Degree | No. |
|---|---|
| Earned astronomy-related doctorates | 37,419 |
| Earned other doctorates | 2288 |
| Highest degree was Master's | 65 |
| Highest degree was Bachelor's | 40 |
| Earned no degree | 4 |
| Highest degree unknown | 3203 |
| Less those entered twice | –55 |
| Total number of persons | 42,964 |

Table 2: Astronomy-related doctorates in AstroGen.

| Details | No. | % |
|---|---|---|
| Astronomy-related doctorates | 37,474 | |
| Thesis titles known | 37,323 | 100 |
| Advisor(s) known | 32,576 | 87 |
| Thesis is in English | 33,481 | 89 |
| Full thesis is online | 20,615 | 55 |
|    Restricted or embargoed | –1026 | –3 |
|    Free to all | 19,589 | 52 |
|    On ProQuest | 5830 | 16 |
| Total available to those with access to ProQuest | 25,419 | 68 |

## 4　HOW CAN YOU USE ASTROGEN?

AstroGen was established for two purposes. The more serious one is to enable studies of the astronomical community and astronomical research by historians and sociologists of science. The other is to allow astronomers the fun of looking up their academic ancestors and descendants. Our role model, the Mathematics Genealogy Project, has been underway since 1996, and it currently contains about 293,000 records. Its Director has found that people also use it for a third purpose: to enable those seeking scientists to referee grant proposals or publications to avoid inviting the students or advisors of the submitters.

　　Since AstroGen is now ten years old, it seems appropriate to provide ten ways you might make use of our project.

### 4.1　Compile a List of Your University's Astronomy Graduates

The most important page on the AstroGen website is the Search page. It allows searching by any combination of name, highest degree, university, thesis title words, years, and country. Choose a university name. If the words in the name are also in the names of other universities, be sure to put the name you want in quotation marks. A search for Sydney yields 322 graduates from five universities. Limiting the search to 'University of Sydney' in quotation marks cuts the number down to 241, including one with a Master's degree and four with non-astronomy-related doctorates. Checking the box for highest degree AD (astronomy-related doctorate), and choosing 'Advanced Search' will yield the 229 records that include all who earned astronomy-related doctorates at the University of Sydney. Click on 'CSV' near the top of the page, and you get a comma separated variable list that you can easily open and manipulate with Microsoft Excel or any other spreadsheet application.

　　Note that you cannot specify a department. Astronomy-related theses are written in Departments of Aerospace Engineering, Atmospheric Science, Chemistry, Computer Science, Geology, Mathematics, Physics, Space Science, and others. Some of the most productive universities have three or more departments where astronomy is done.

### 4.2　Find the Theses Based on Observations Made at Your Observatory

Observatories track publications, including dissertations (theses), which make use of data from their instruments. While the SAO/NASA Astrophysics Data System (ADS) allows curators to readily find journal articles of interest, the coverage of dissertations continues to be difficult. The aggregation of links to dissertations which the Astronomy Genealogy Project provides makes this an important resource for finding observatory-related dissertations.

　　For example, the Chandra X-ray Center Bibliography lists 1114 Ph.D. theses that include Chandra data (https://cxc.harvard.edu/cgi-gen/cda/bibliography). Sherry Winkelman, then Chief Archive Specialist at the Chandra X-ray Center, compiled most of this list and showed that a substantial portion of it came from AstroGen. She first started with a list of dissertations that were known to incorporate Chandra data, then followed up and down the family trees of these astronomers to check the theses of their advisors and students (Winkelman and Tenn, 2021).

### 4.3　Compare the Production of Astronomy Related Doctorates by Country

Currently, we claim that AstroGen is 'nearly complete' for 38 countries and other polities, although some of them have not been updated for a year or two or three. We chose to complete these first because the volunteers we have are knowledgeable about their languages and academic practices. We will need other volunteers to compile data from countries with different alphabets and practices. We can't find our way through a library website when the site uses an unfamiliar alphabet, even if the theses are in English.

　　There is a table on our website, continuously updated, listing these countries. It is in the FAQs, under "How complete is AstroGen?" Table 3 shows its current contents. Although the online table can be sorted by any column, the default is to order the countries by the number of astronomy-related doctorates per million residents. Unsurprisingly, the countries that rank highest on this criterion are the ones that attract the most international students. The United Kingdom and the Netherlands, both of which attract a great many students from abroad, have each produced more than 70 astronomy-related Ph.D.s per million population.

There are several countries that produce a lot of doctorates but are not yet complete in AstroGen. We have compiled more than 2700 degrees from universities in France, but most are not yet online. We are waiting for a knowledgeable person to help us sort out the changes in French universities, which have been on a merging binge in recent years.

Table 3: Countries deemed 'nearly complete'.

| Country | Astronomy Related Doctorates | Population in millions (UN, 2022) | Deg's per million |
|---|---|---|---|
| United Kingdom | 5663 | 67.5 | 83.9 |
| Netherlands | 1271 | 17.6 | 72.2 |
| Switzerland | 472 | 8.7 | 54.3 |
| United States | 17,177 | 338.3 | 50.8 |
| Germany | 3947 | 83.4 | 47.3 |
| Australia | 1214 | 26.2 | 46.3 |
| Sweden | 474 | 10.5 | 45.1 |
| Finland | 237 | 5.5 | 43.1 |
| Spain | 1468 | 47.6 | 30.8 |
| Canada | 1164 | 38.5 | 30.2 |
| Belgium | 351 | 11.7 | 30.0 |
| Estonia | 39 | 1.3 | 30.0 |
| Denmark | 160 | 5.9 | 27.1 |
| Ireland | 132 | 5.0 | 26.4 |
| Greece | 273 | 10.4 | 26.3 |
| Israel | 226 | 9.0 | 25.1 |
| New Zealand | 114 | 5.2 | 21.9 |
| Austria | 144 | 8.9 | 16.2 |
| Norway | 83 | 5.4 | 15.4 |
| Lithuania | 40 | 2.8 | 14.3 |
| Serbia | 72 | 7.2 | 10.0 |
| Argentina | 353 | 45.5 | 7.8 |
| Iceland | 3 | 0.4 | 7.5 |
| Portugal | 61 | 10.3 | 5.9 |
| Chile | 108 | 19.6 | 5.5 |
| Croatia | 22 | 4.0 | 5.5 |
| Mauritius | 5 | 1.3 | 3.8 |
| Azerbaijan | 34 | 10.4 | 3.3 |
| South Africa | 178 | 59.9 | 3.0 |
| Türkiye | 227 | 85.3 | 2.7 |
| Brazil | 536 | 215.3 | 2.5 |
| Hong Kong | 13 | 7.5 | 1.7 |
| Singapore | 3 | 6.0 | 0.5 |
| Iran | 30 | 88.6 | 0.3 |
| Columbia | 6 | 51.9 | 0.1 |
| Pakistan | 15 | 235.8 | 0.1 |
| Nigeria | 10 | 218.5 | 0.0 |
| Ethiopia | 5 | 123.4 | 0.0 |
| Totals | 36,330 | 1900.3 | 19.1 |

It is worth pointing out that universities in these 38 countries have produced 97% of all the doctorates currently in our database. Here is a question for the reader: What do you think is the median year for the award of all these astronomy-related doctorates? The answer will appear somewhere below.

### 4.4 Compare the Production of Astronomy Related Doctorates by University

AstroGen goes back to the beginnings of the modern, research-based doctorate, now called a Doctor of Philosophy degree in most countries and abbreviated Ph.D. in all but a few universities. We are somewhat incomplete on the earliest such degrees, which were awarded in the eighteenth and nineteenth centuries (we need a Latin scholar), but the number to be added is undoubtedly small. Figure 4 shows the cover of the oldest thesis currently in Astro-Gen. We are not certain that it fits our criteria, but the Latin scholar who reviewed it for us reported that it "… is a sort of prospectus for a larger work …", discussing what would be needed "… to develop a method for precisely predicting the phases of Saturn's ring." (R. Ceragioli, pers. comm., 19 February 2023). The twenty-five most productive universities in our database are listed in Table 4. All now accept all or nearly all their theses in English, and most of them attract many students from other countries.

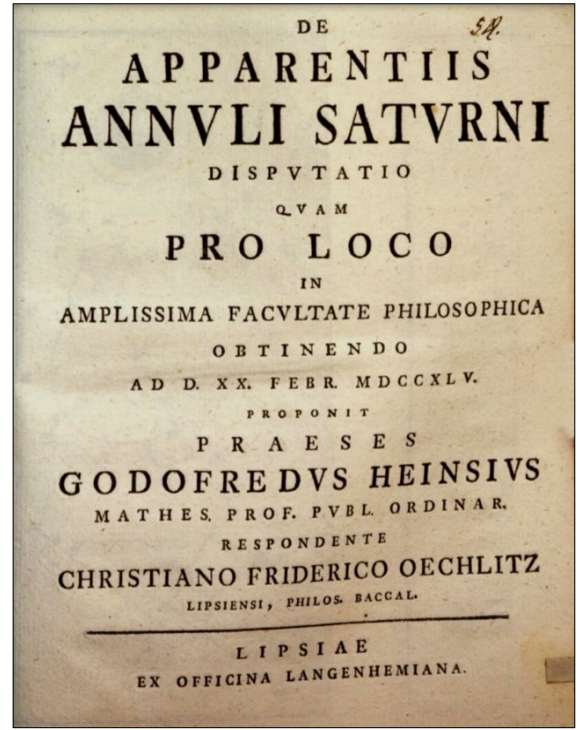

Figure 4: The cover of the oldest thesis currently in AstroGen. Christian Friedrich Oechlitz (dates unknown) submitted this thesis on the rings of Saturn to Leipzig University in 1745, where Gottfried Heinsius (1709–1769) was his advisor. Like many theses that are out of copyright, it is freely available online, in this case on the site of the Bavarian State Library (https://www.digitale-sammlungen.de/view/bsb10660167?page=1).

### 4.5 Compare the Careers of Graduates of Different Universities

I spent more than thirty years advising undergraduate physics majors. One topic that came up frequently was how to choose a graduate school. If you are an advisor to students applying to doctoral programs in astronomy, you might find it useful to use AstroGen to compare the careers of astronomy Ph.D.s from different universities. This takes a bit of work, but AstroGen can get you started.

Table 4: Universities (currently 'nearly complete' in AstroGen) that have produced the most astronomy-related doctorates.

| University | Theses |
|---|---|
| University of California, Berkeley | 781 |
| University of Cambridge | 765 |
| California Institute of Technology | 748 |
| Heidelberg University | 727 |
| University of Arizona | 641 |
| Harvard University | 605 |
| University of Manchester | 550 |
| University College London | 534 |
| University of Chicago | 527 |
| LMU Munich | 522 |
| Massachusetts Institute of Technology | 481 |
| University of Maryland, College Pk | 479 |
| University of Michigan | 476 |
| University of Colorado Boulder | 475 |
| Princeton University | 468 |
| University of Texas at Austin | 458 |
| University of California, Los Angeles | 454 |
| Leiden University | 442 |
| Cornell University | 415 |
| University of Wisconsin-Madison | 384 |
| University of Bonn | 361 |
| University of Oxford | 360 |
| University of La Laguna | 343 |
| University of Leicester | 331 |
| Columbia University | 330 |

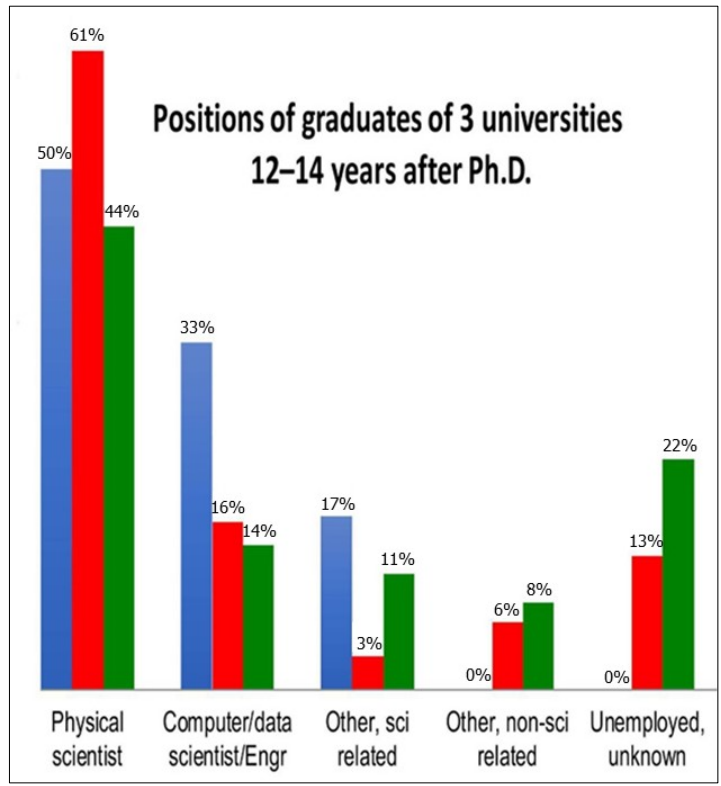

Figure 5: Positions held by astronomy graduates of three highly-productive universities 12–14 years after graduation. Note that they earned their doctorates in 2006–2008, and positions are as of 2020.

As an example, we consider the careers of graduates of three (anonymous) universities from among the 25 most productive ones. Each had 30 to 36 Ph.D. graduates in the years 2006–2008. AstroGen makes it easy to download lists of names and degrees and to follow links to personal web pages of the graduates to learn what they are doing currently. This was done in October 2020, twelve to fourteen years after graduation. By this time most were no longer moving from one postdoctoral position to another every two to three years. We found the results shown in Figure 5. A student whose goal is to obtain a permanent position as a research scientist might choose the university portrayed in red, while one who aims to become a data scientist or engineer might prefer the university shown in blue.

### 4.6 Trace Changes in the Language Used for Writing Theses

Practically all doctoral theses were written in Latin until the late nineteenth century. Most scholars then switched to the language of their university. Starting early in the twentieth century, one country after another switched to writing scientific theses in English, until by the beginning of the twenty-first a majority of astronomy-related theses were being written in English in many countries. Table 5 shows some examples. The transitions in Germany were particularly striking, as can be seen in Figure 6. The points represent ten-year bins, centered on the year shown. The two English points at the decades centered on 1860 and 1870 each represent one American who was allowed to write his thesis in English at the University of Göttingen.

The trend to a single common language has accompanied the globalization of many graduate programs. Figure 7 shows countries in which at least one university offers the possibility of earning an astronomy-related Ph.D. without knowing any language but English. A particularly cosmopolitan doctoral program is shown in Table 6. Only one-fourth of those who earned astronomy-related doctorates at the Leiden University in 2021 and 2022 were from the Netherlands. They were valuable to their classmates, as they were often called upon to translate thesis abstracts into Dutch, as re-quired by the University, but it seems clear that seminars and conversations in the astronomy program were held in English, the language in which all astronomy theses at Leiden have been written since 1964.

Table 5: Theses in English for selected countries.

| Country | Majority in English Since | All in English Since | % in English (1850–2021) |
|---|---|---|---|
| Netherlands | 1953 | 1964 | 96 |
| Finland | 1967 | 1967 | 99 |
| Denmark | 1979 | 1979 | 97 |
| Sweden | 1984 | 1994 | 97 |
| Israel | 1998 | 2009 | 87 |
| Switzerland | 1998 | — | 81 |
| Austria | 2000 | 2019 | 79 |
| Germany | 2002 | — | 72 |
| Spain | 2010 | — | 50 |
| Chile | 2011 | 2020 | 79 |
| Greece | 2017 | — | 17 |

### 4.7 Trace the Growth and Decline of Different Fields of Astronomy

A little over a century ago, a typical astronomy-related doctoral thesis consisted of measuring the positions of a comet or asteroid or using someone's observations to calculate its orbit. Some astronomers did both. Decades later, there was a time when nuclear astrophysics and construction of stellar models were among the most popular topics. Currently, it seems to be exoplanets. AstroGen provides a relatively easy way to quantify such changes by searching on title words over time. Figure 8 shows some examples. Both 'pulsar' and 'quasar' start-ed appearing in thesis titles in the five-year period centered on 1970, and both rose rapidly. Theses on both have decreased a bit in the last ten to fifteen years, as the number on exoplanets has soared. Note that I have combined 'extrasolar planet' with 'exoplanet', as the two terms have the same meaning.

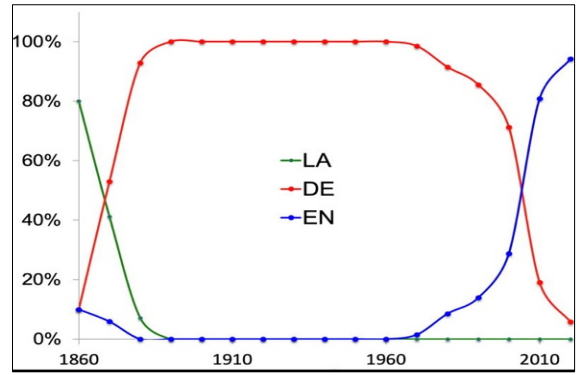

Figure 6: The languages in which astronomy-related doctoral theses were written in Germany. German replaced Latin in the late nineteenth century, and English became dominant near the beginning of the twenty-first century.

Table 6: Leiden Ph.D.s 2021–2022 by country of birth.

| Country | Number |
|---|---|
| Netherlands | 8 |
| Italy | 3 |
| United States | 3 |
| France | 2 |
| Germany | 2 |
| Poland | 2 |
| United Kingdom | 2 |
| Belgium | 1 |
| Brazil | 1 |
| Chile | 1 |
| China | 1 |
| Estonia | 1 |
| India | 1 |
| Ireland | 1 |
| Spain | 1 |
| Türkiye | 1 |
| Ukraine | 1 |
| Total | 32 |

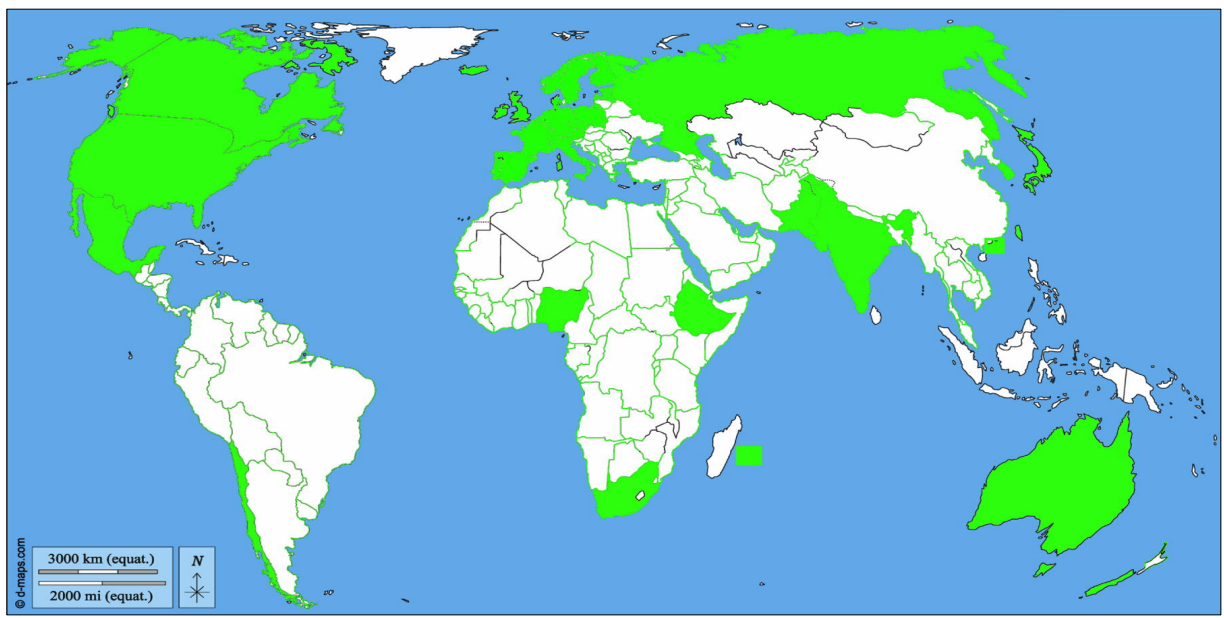

Figure 7: World map with color showing countries in which at least one university offers an astronomy-related doctorate with (post-master's) graduate instruction and thesis entirely in English.

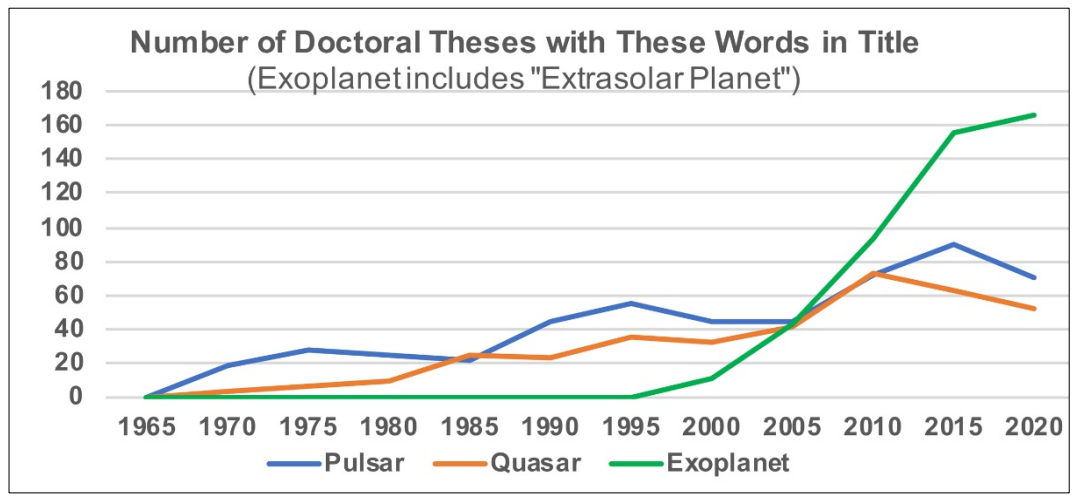

Figure 8: Changes over time in popularity of some fields of astronomy. Points represent five-year bins centered on the number on the horizontal axis.

### 4.8 See How the Progeny of Prize-Winning Astronomers Fare

I looked up the recipients of nine major awards over the past century (1923–2022). For each award I included only those who received them for astronomy-related research. A summary of the results is in Table 7. A whopping 32% of award recipients were students of award recipients. If we only consider recipients with earned doctorates, which is now nearly universal, then it is 35%.

Figure 9 is a family tree of one awardee, showing how many of his academic ancestors had been recognized with comparable prizes.

Table 7: Recipients of nine major awards in 1923–2022.

| Award | Since | No. |
|---|---|---|
| Gold Medal, RAS (G) | <1923 | 125 |
| Draper Medal, NAS (D) | <1923 | 38 |
| Bruce Medal, ASP (B) | <1923 | 98 |
| Nobel Prize (N) | <1923 | 26 |
| Russell Lectureship, AAS (R) | 1946 | 75 |
| Crafoord Prize (C) | 1985 | 13 |
| Gruber Cosmology Prize (GP) | 2000 | 43 |
| Shaw Prize (S) | 2004 | 31 |
| Kavli Prize (K) | 2008 | 19 |
| Total Awards | | 468 |
| Correction for multiple recipients | | −207 |
| Number of award-winning individuals | | 261 |
| Number of earned doctorates | | 240 |
| Number of thesis advisors who received at least one of these awards | | 83 |

### 4.9 Find Academic Ancestors of an Astronomer

AstroGen provides ancestral 'family trees' for almost everyone with an astronomy-related doctorate. Figures 9 and 10 are examples.

### 4.10 Find Academic Descendants of an Astronomer

This is more difficult, as an astronomer's page shows only his or her students and not their academic progeny. You must click on each one separately to see that student's students, and repeat to the end. We have not yet achieved the ability to plot family trees of descendants, but we do provide one generation of descendants as the default option with the ancestral family trees. Figure 10 shows the academic ancestors and children of an astronomer who was born and raised in Zimbabwe, earned her doctorate in the United Kingdom, and now has her own doctoral students in South Africa.

### 5 WHAT'S NEXT?

Although ten years old, AstroGen is still very much a work in progress. We have yet to obtain much information about theses submitted to universities in China, Italy, Mexico, Russia, and many other countries, most of them in Asia. We have some theses from Japan and Korea, but we need many more. We need volunteers who know the languages and can find their way through the library websites of universities in these countries. In India, the theses are in English, but we need a volunteer familiar with the names and the academic culture to add to the 326 theses currently in our database. The earliest research-based Ph.D. degrees constitute another area where help is needed, in the form of a volunteer who can translate Latin titles and, preferably, has some familiarity with nine-teenth century astronomy. Eventually, we hope to get volunteers to go into university libraries and examine the theses that are not online. And for those who have been waiting to find the answer to the question posed above, one-half of the doctorates currently listed in AstroGen have been awarded since 2004. If we restrict the count to astronomy-related theses, then the median year is 2006.

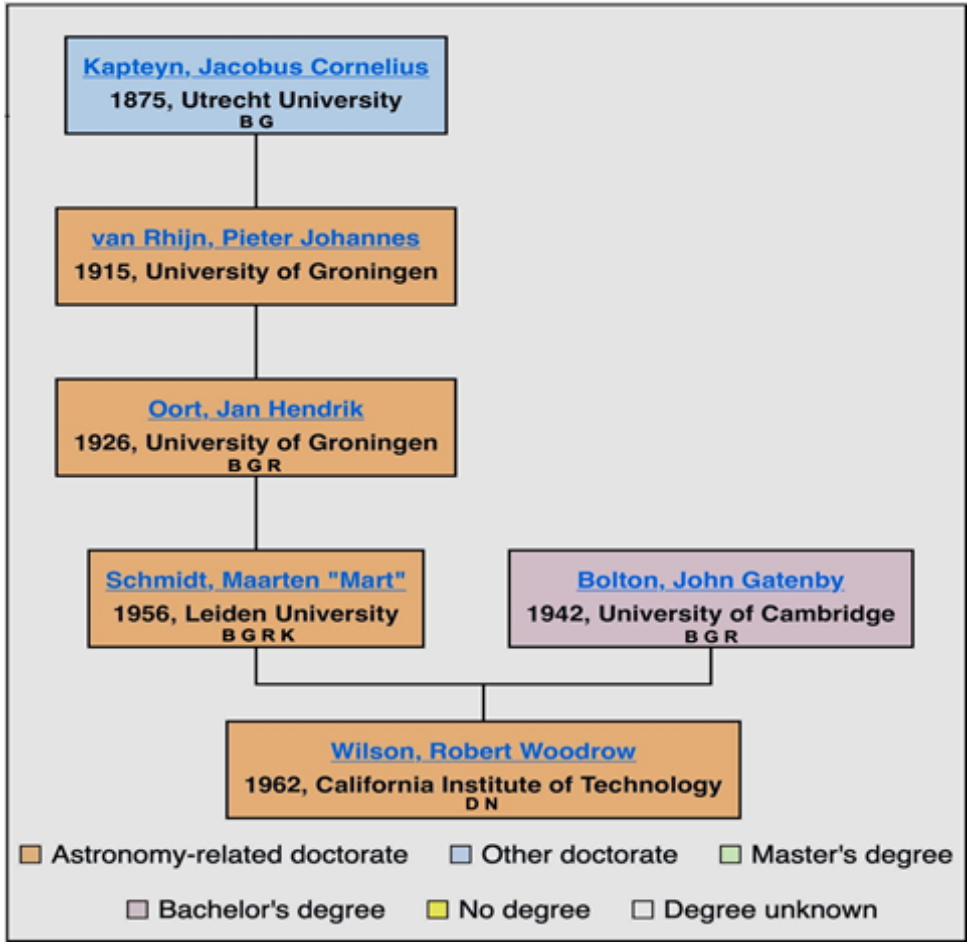

Figure 9: The ancestral family tree of Robert Woodrow Wilson. Letters below the University name have been added to indicate the awards received, as detailed in Table 7.

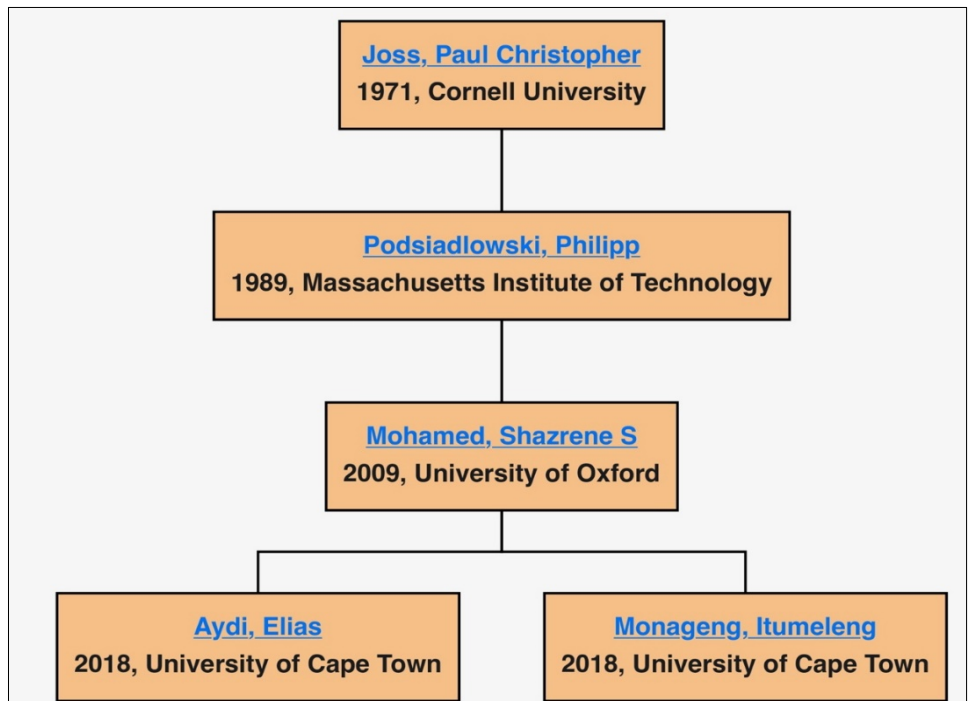

Figure 10: The family tree of Shazrene Mohamed, a native of Zimbabwe who earned her doctorate in the United Kingdom, showing the doctoral students she has supervised in South Africa.

## 6 ACKNOWLEDGMENTS

There is a full-page list of acknowledgments on the AstroGen website, listing many individuals who have made large or small contributions to the project. Here I will mention only a few of the most important. The AAS funds the programming and hosts AstroGen on its website. Jonathan Galantine is the ace programmer who rescued the project when it was floundering. We received valuable advice on building the website from Mitch Keller (Mathematics Genealogy Project), Arnold Rots (Center for Astrophysics | Harvard & Smithsonian), and Kenneth Ritley (Bern University of Applied Sciences). Major data contributors have included Younes Ataiiyan (Santa Rosa Jr. College), Jennifer Lynn Bartlett (US Naval Academy), R. Peter Broughton, Vassilis Charmandaris (University of Crete), Wolfgang R. Dick, James Lattis (University of Wisconsin-Madison), Tsuko Nakamura (Daito Bunka University), Wayne H. Osborn (Central Michigan University), Selim O. Selam (Ankara University), Horace Smith (Michigan State University), Benny Trakhtenbrot, (Tel Aviv University), Carlos Viscasillas Vázquez (Vilnius University), and Michael J. Way (NASA/Goddard Institute for Space Studies).

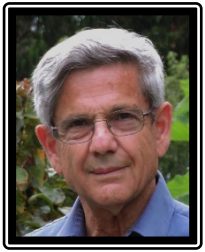

**Professor Joseph S. Tenn** taught Physics and Astronomy at Sonoma State University in the California wine country from 1970 to 2009.

He served as Secretary-Treasurer of the Historical Astronomy Division of the American Astronomical Society from 2007 to 2015 and as an Associate Editor of the *JAHH* from 2008 to 2016. He maintains the ASP Bruce Medalists website at http://phys-astro.sonoma.edu/brucemedalists.

He has devoted the past decade to directing the Astronomy Genealogy Project, which he founded. His ORCID number is 0000-0002-7803-3633.